# Leveraging Cognitive Search Patterns to Enhance Automated Natural Language Retrieval Performance

B. Selvaretnam
Faculty of Computer Science
Multimedia University
selvaretnam@mmu.edu.my

M. Belkhatir
Faculty of Computer Science
University of Lyon
belkhatir@univ-lyon1.fr

**ABSTRACT**
The search of information in large text repositories has been plagued by the so-called *document-query vocabulary gap*, i.e. the semantic discordance between the contents in the stored document entities on the one hand and the human query on the other hand. Over the past two decades, a significant body of works has advanced technical retrieval prowess while several studies have shed light on issues pertaining to human search behavior. We believe that these efforts should be conjoined, in the sense that automated retrieval systems have to fully emulate human search behavior and thus consider the procedure according to which users incrementally enhance their initial query. To this end, cognitive reformulation patterns that mimic user search behaviour are highlighted and enhancement terms which are statistically collocated with or lexical-semantically related to the original terms adopted in the retrieval process. We formalize the application of these patterns by considering a query conceptual representation and introducing a set of operations allowing to operate modifications on the initial query. A genetic algorithm-based weighting process allows placing emphasis on terms according to their conceptual role-type. An experimental evaluation on real-world datasets against relevance, language, conceptual and knowledge-based models is conducted. We also show, when compared to language and relevance models, a better performance in terms of mean average precision than a word embedding-based model instantiation.

**Keywords**
Information Search, Query Enhancement, Linguistic Knowledge Acquisition, Human Search Patterns, Genetic Algorithms

## 1. INTRODUCTION & RELATED WORKS

To address the mismatch issue between query and document vocabularies brought about by the ambiguity of natural language, several technological solutions have consisted in incorporating linguistic, statistical and semantic-based techniques in the information retrieval process. Parallel to these works, a community has emerged aiming at studying the intricacies and consequences of human behavior while searching for information. Brought about by the findings obtained by these research efforts, the need to enhance information retrieval performance through the comprehension of the intended human search goal by considering query components has been highlighted. As such, query enhancement frameworks often rely on lexical resources so as to highlight novel terms which are semantically related to the original query entities and perform sense disambiguation. Query elements inherently possess two different functionalities which are often not taken into account: concepts are either utilized to connect query components or characterize the intent by translating the user's search goal. It is therefore important to consider the fact that query constituents can play distinct roles that, if determined and considered appropriately would yield a more sensible comprehension of the conveyed user's search goal. Furthermore, the intricacies of human language impose that there are relations between neighboring and non-adjacent concepts stressing semantic characterizations intrinsically attached to the search goal. Earlier propositions mostly do not succeed in fully capitalizing on the relations between query constituents, which, if used suitably, would improve retrieval effectiveness. As a matter of fact, adjacent and non-adjacent dependencies can be captured through full dependence query representation. However, this process will prove computationally costly in especially longer queries as multiple concept pairs will be produced, among which a large number would possible be deemed meaningless. Indeed, the process of query enhancement based on these pairs would generate non-related concepts, and consequently cause digression from the user's search goal. We formulate the hypothesis that neighboring and distant associations between query terms can be shown through syntactic dependencies. This in turn translates into relevant concept pairing for query enhancement.

Additionally, instead of making use of statistical processes for producing extra concepts, the use of linguistic resources such as ontologies make it possible to spawn then encapsulate in the original query enhancement concepts that are semantically relevant with its content. These solutions appear even more attractive as the most recent systems based on word embeddings (Roy et al. 2016) fail to achieve retrieval performance on par with statistical-based techniques. The effectiveness of extrinsic knowledge-based enrichment is however highly dependent on the successful disambiguation of query concepts. In general, the modest results of semantic disambiguation methods have resulted in techniques attempting to improve its accuracy with the utilization of information extracted from external sources (e.g. Mihalcea, 2007), improved relatedness measures (e.g. Patwardhan et al., 2007) or deep learning (Festag & Spreckelsen, 2017). On the basis of the derived sense of query terms, both lexical-related and semantically linked enhancement concepts are determined from a given ontological resource. Terminologies in domain-related ontologies are less prone to ambiguity, consequently short queries can be enhanced with greater effectiveness. Generic ontologies would themselves be suitable for broader queries (Bhogal et al., 2007). In previous solutions dealing with external knowledge sources, Liu et al. (2004), Chauhan et al. (2012) utilize generic ontologies; Alejandra Segura et al. (2011), Bhogal et al. (2013) use domain-related ontologies while Tuominen et al. (2009) evaluate their framework with both general and domain-related ontologies.

Having discussed related techniques, it is observed that both statistical processing and extrinsic linguistic based resources bring about benefits to the query enhancement process. Knowledge based processing allows for lexical-semantic related terms to be obtained while statistical processing allows for statistically-collocated terms to be extracted to ensure that all relevant documents can be retrieved for a given query. Due to the data sparsity problem, it has been shown that, even with large amounts of data, significant evidence of potential term relationships is not fully captured. Consequently, our aim of attempting to reconcile multiple sources of evidence, i.e. linguistic, statistical and extrinsic knowledge resource-based, in order to predict query intent more accurately is a promising paradigm.

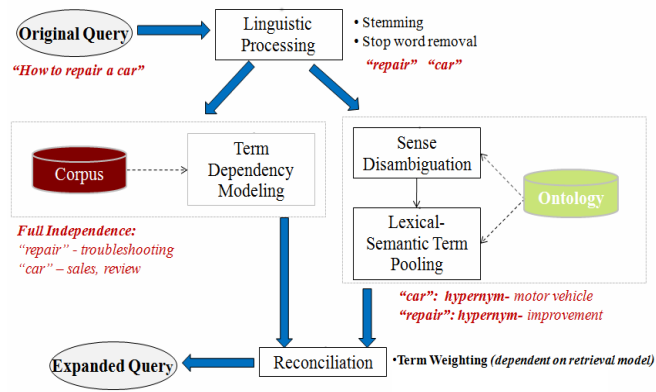

**Figure 1: Process Flow of an Integrated Query Enhancement Framework**

In figure 1, the process flow of such an integrated query enhancement framework is depicted. An original query is subjected to fundamental linguistic processing of stemming and stopword removal. The remaining terms serve as input for term dependency modelling in statistical and ontology-based processing to extract expansion concepts. The extracted concepts are reconciled via term weighting models that are embedded within an existing retrieval model. Accordingly, a capital issue in query enhancement is the procedure of weighting the original and enhancement concepts to reflect the search goal adequately. Until now, related works have emphasized the notion of document-based concept frequency by means of calculating document and term frequency (e.g. Song et al., 2008) or the development of supervised learning techniques featuring several variables and also considering the computation of concept frequency considering distinct sources such as query logs and n-grams (Paik & Oard, 2014). The issue arising with such solutions relies on the fact that head concepts are determined according to statistical occurrence, which is not de facto indicative of the search objective. Alternatively, it is our postulation that concepts should be awarded due importance with respect to the conceptual role they play in representing the information need.

After presenting and analyzing integrated query enhancement and the related approaches in Section 2, we introduce the conceptual instantiation of a query in Section 3 and in Section 4 propose the formalization of the retrieval model. Components of the proposed architecture are detailed in Section 5 and an algorithmic characterization presented in Section 6. An empirical evaluation of the proposed framework on the TREC 1, 3, 8 and Terabyte datasets is described in Section 7.

## 2. INTEGRATED QUERY ENHANCEMENT

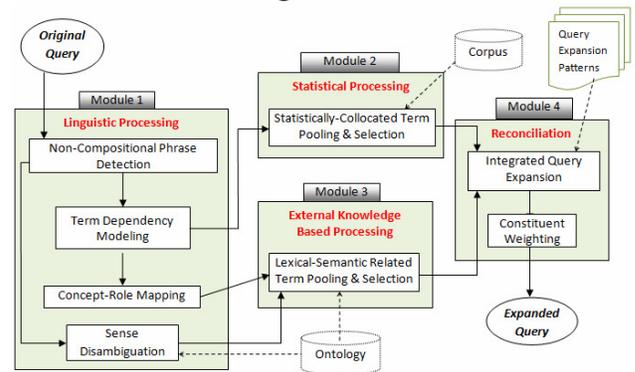

**Figure 2: Anatomy of a Query Enhancement Framework Integrating Statistical and External Linguistic Knowledge Based Processing**

The anatomy of a query expansion framework integrating statistical and external linguistic knowledge based processing which consists of four modules is presented in Figure 2. The intent of the user's search goal is inferred and encoded through linguistic analysis (**Module 1**); related terms are extracted from multiple sources in accordance to the types of query-document links that are identified (**Module 2 & Module 3**); and finally an expanded query is formed by integrating the pooled candidate expansion terms via specific query expansion patterns and weighting scheme (**Module 4**).

At the core of this framework are the significant constituents that characterize the query intent. Their identification within a query is crucial to determine only terms which directly express the information need of the user and are not merely structural components of a query. In this article, intent is defined based on Frege's Principle of Compositionality (Partee, 1995). The Principle of Compositionality is defined as "*the meaning of a whole is a*

*function of the meanings of the parts and of the way they are syntactically combined"*. Specifically, the meaning of a sentence is determined by the structure of the sentence and the meanings of the constituents within the sentence (Szabó, 2008). Thus, query intent is reflected by its constituents' roles which are identified by recognizing the syntactical structures within a query.

Syntactical structures are obtained by parsing a query in accordance to Chomsky's transformational grammar (Chomsky, 1957). The ordered, rooted non-binary parse tree obtained is representative of the syntax of a query in the form of syntactic categories i.e. phrases (e.g. noun phrase, verb phrase etc) as well as words labelled according to their parts-of-speech (POS). However, prior to syntactical parsing, non-compositional phrases (NCP) are isolated to retain the compounded term in the intended form. Apart from that, grammatical dependencies (Marneffe et al., 2006) which are represented as directed graphs exist within queries and allow for global inter-relationships between query terms to be inferred. In the case of polysemic terms, these are disambiguated either through supervised/unsupervised methods which are reliant on knowledge sources (e.g. text corpora, dictionaries etc) to associate the most appropriate sense to a term in context (Navigli, 2012). The linguistic processes for determining the significant query constituents and their role types are handled in **Module 1**.

To bridge the query-document vocabulary mismatch problem, *related* expansion terms are added on to an original query. ***Related terms*** are defined as the set of terms which are directly or indirectly associated to an original query term. A directly related term is one which is of an exact match to the original term. Indirectly related terms are those which frequently occur within close proximity (adjacent and/or non-adjacent) as well as terms which share a lexical-semantic link. Lexical-relations which are embodied by synonymous term relationships may be classified in a flat hierarchy whilst semantic links are usually organized in parent-child tree structures which indicate generalization (e.g. hypernyms) or specialization (e.g. hyponyms). The vocabulary of **Relevant Documents** would contain such related terms as the central topic of a document is often described with a combination of general and specific terms (Chemudugunta & Steyvers, 2007). Thus, at a local level, the query constituents are subjected to statistical processing where term dependency is modelled for statistically-collocated term pooling (Huston&Croft, 2014) from a document corpus/n-gram model. External knowledge based processing is performed where semantic ambiguity is resolved and subsequently lexical-semantic related terms from external knowledge sources i.e. ontologies are pooled (Selvaretnam&Belkhatir, 2019). The expansion term pooling processes involving term selection from appropriate knowledge sources is tackled in **Modules 2 & 3**.

Reformulation of a query is a common occurrence when a search task is performed (Hearst, 2009). Reformulation takes place either due to unsatisfactory results obtained from an initial search or a change in information need. Inaccurate search results which are due to a mismatch in query-document vocabulary are rectifiable in an ad-hoc retrieval process. However, emulation of a user's search pattern or preference is of the essence in ensuring effective retrieval that is in line with the original information need (Jansen et al., 2008; Rieh & Xie, 2006). Proximity and lexical-semantic related candidate expansion terms require filtering to ensure that the degree of specificity or generalization of a query does not cause extensive deviation from the original query intent through expansion. Further, each term within the expanded query should be assigned suitable weightage on their own accord. The integration of potential expansion terms derived from statistical and external knowledge-based sources in accordance to specific query expansion patterns and appropriate weighting of the query constituents is handled in **Module 4**.

## 3. CONCEPTUAL QUERY MODELING

We propose to characterize a query through a set of concepts rather than considering a string representation or a bag-of-words characterization. At the core of our proposal is the notion of Concepts-of-Interest (*CoIs*), introduced as the head concepts therefore reflective of search intent. We furthermore heuristically argue that query components are classified into four supplementary categories (or *role-type*s) translating their intrinsic role:

- Descriptive Concepts (*DCs*) provide further descriptive details related to head concepts;
- Relational Concepts (*RCs*) highlight a relationship between query concepts;
- Structural Concepts (*SCs*) are stopwords that assist in forming the structure of a query;
- Enhancement Concepts (*ECs*) are produced in order to further enrich the initial query.

Procedures to highlight these different concepts will be presented in Section 5 while a complete algorithmic instantiation will be provided in Section 6.

The query conceptual representation is defined by the triplet $<C_q, R_q, A_q>$ where:

- $C_q$ is the set including *ECs*, *RCs*, *DCs* and *CoIs*.
- $R_q$ is the set of structural, syntactical and semantic relations.
- $A_q$ is the set including axioms defined over $C_q$ and $R_q$.

We shall now consider the example natural language query "How to repair a car with engine failure".

The conceptual assignation process presented in Section 5.1 tags verbs and nouns as *CoIs* and the remaining parts-of-speech constitute the structural relations and concepts.

The valid sense for each *CoI*, i.e. 'engine failure', 'car' and 'repair', is then highlighted. A derived semantic relation stresses that *car* is a holonym of *engine*. Making use of this information, the related semantic concepts are generated. Considering the *CoI* 'repair', synonyms only are considered since it is a verb. As far as the collocation 'engine failure' is concerned, coordinate terms, hypernyms, meronyms, synonyms are pooled and only coordinate terms, hypernyms and synonyms for the *CoI* 'car'. The procedure of conceptual enhancement is developed in Section 5.2.

The resulting conceptual representation of the example query is presented in Figure 3.

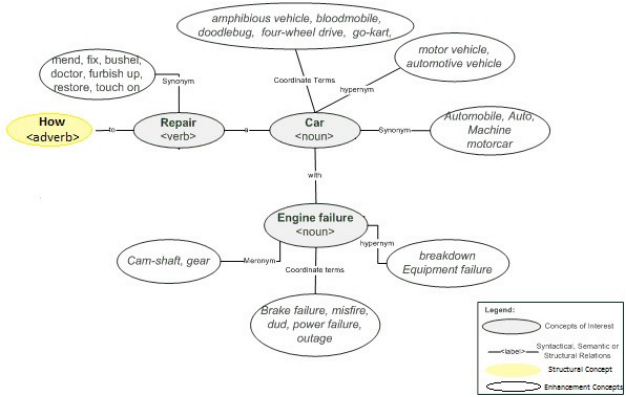

**Figure 3: Generation of the conceptual structure for the example query "How to repair a car with engine failure"**

## 4. RETRIEVAL MODEL INSTANTIATION

The considered retrieval model computes the relevance *Rel(d,q)* of an index document *d* with respect to a query *q* as follows :

$$Rel(d,q) = f\,(Sat(d,q)\,,\,Imp(q,d))$$

While function *Imp* allows quantifying the importance of the concepts of *q* in *d*, function *Sat* is a measure of the extent to which *d* satisfies *q*. Both consider only documents that comprise the query concepts, i.e. *CoIs*, *DCs*, *RCs*, *SCs* and *ECs*, fully or partially. As far as the *f* function is concerned, it produces values proportional to the values of the *Sat* and *Imp* functions through the trivial multiplication operator.

### 4.1 Formalization of the *Sat* function

Function *Sat* considers a language model $L_d$ with respect to *d* and then quantifies to which extent it satisfies *q* by computing the log-likelihood of producing *q* given $L_d$ with *Sat(d,q)*:

$$Sat(d,q) = log\ p(q|L_d)$$

$L_d$ is chosen to be a multinomial distribution according to the state-of-the-art, i.e. $p(q|L_d) = \Pi_c\ p(c|L_d)^{N_q(c)}$ (where $N_q$ computes the number of times concept *c* appears in *q*), and with maximum likelihood estimator and the use of Dirichlet prior smoothing to address the zero-probability issue, we obtain:

$$p(c|L_d) = \frac{N_d(c) + \mu.N_c(c).|C|^{-1}}{|d| + \mu}$$

where the $N_d$ and $N_c$ functions count the number of times a given concept appear in a document and the whole collection respectively; $|d|$ and $|C|$ are the number of concepts in the document and the collection respectively; $\mu$, the smoothing parameter, is a constant.

### 4.2 Formalization of the *Imp* function

Function *Imp* considers the importance of the concepts of *q* within *d*. We indeed postulate that the user expects, upon formulating his query, that the documents retrieved convey the query topics. If additional themes which are not formulated in the query are present in the document, then they shall be tightly semantically related to the initial query concepts. For the example query "How to repair a car with engine failure", documents mentioning 'motorcars' (i.e. a synonym of 'car') will be considered relevant while documents with 'boats' will not. This concept of 'semantic closeness' is an important factor guiding the development of the enhancement process and the spawning of *ECs*.

Here, the impact value *Imp(q,d)* measures the importance of the semantic themes of query *q* within the document *d* by considering the weight of the query concepts and their number (i.e. $|q|$):

$$Imp(q,d) = \Sigma_{sc\ semantic\ concept\ of\ q\ corresponding\ to\ a\ concept\ of\ d}\ W(sc).|q|^{-1}$$

Weight function *W* quantifies the importance of query concepts as their impact is not uniquely determined by their functionality. Search performance is significantly improved by the incorporation of significant concepts such that the shared theme vehiculated by the global pool of concepts strongly impacts retrieval performance. We may intuitively assume that *CoIs* and *DCs* hold a more important role than *SCs*, *RCs* and *ECs*. This assumption may however not necessarily hold true depending on the shared theme of query concepts. Therefore a procedure allowing to progressively adapt the concept weights to optimize query performance is formulated based on a Genetic Algorithm (GA). GA is used for the estimation of role-type weights considering a query for which the measure of search performance is optimized. The optimization process is based on three processing steps: (i) Determination of a population of chromosomes, structures encoding potential solutions through the characterization of five genes representing each role-type; (ii) Definition of a fitness function to evaluate each solution according to the performance metric; (iii) Selection of the fittest individual (i.e. for which the metric is maximal) further used to reproduce. The optimum rates and parameters utilized in the GA implementation are derived from running a tuning step on fifty queries (i.e. test cases) until reaching a stabilized state corresponding to optimum retrieval performance. The evolution of the population of chromosomes is performed according to an optimal iteration count as a terminating condition. Perturbations to the chromosomal population are randomly inflicted through mutation and cross-over.

# 5. DESCRIPTION OF THE QUERY PROCESSING FRAMEWORK & COMPONENTS

Fig. 4 depicts an overview of the stages of processing the example query "coping with overcrowded prisons" will be subjected to in the query enhancement framework. Block 1 is based on the application of linguistic techniques followed by two separate modules of statistical processing on the one hand and extrinsic knowledge-based processing on the other hand to generate expansion terms. Please note that the principles employed here have been extensively covered in the literature (Selvaretnam&Belkhatir, 2012) and will be only summarized in Section 5.1. The core of our current proposal detailed in Section 5.2 is encapsulated within the Integrated Query Enhancement Block (2) where the previously generated expansion terms are selected and reconciled according to reformulation patterns and weighted based on their role-type.

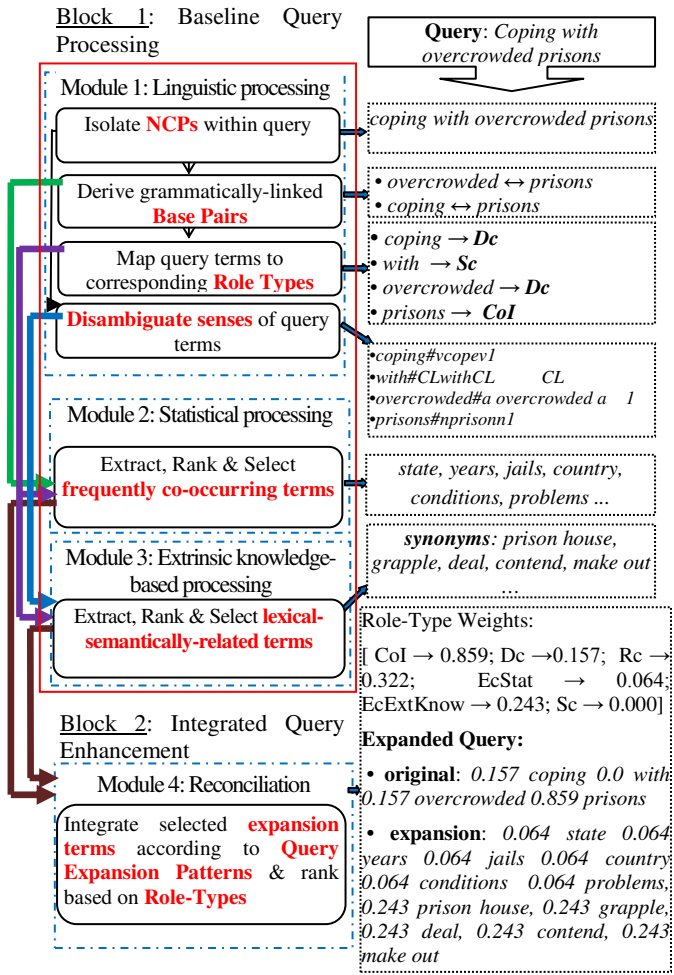

**Figure 4: Processing flow of query "coping with overcrowded prisons"**

## 5.1 Baseline Concept-based Query Processing

The first task consists in identifying non-compositional phrases (NCPs) so as to preserve the semantic intent of the user query. Among the various types of NcPs within the English language, idioms, phrasal verbs, collocations, acronyms and proper names are especially noteworthy as they provide a direct contribution towards the search goal.

Then, the characterization of a natural language query under the prism of lexical-semantic analysis determines constituents that play the role of either function or content components. Content words (e.g. verbs, nouns, adverbs and adjectives) refer to objects, actions and ideas in the everyday world. They can also possibly serve as modifiers and complements that enrich the meaning of a query. Function words (e.g. determiners, pronouns, auxiliaries, prepositions, conjunctions), on the other hand, express grammatical features such as definiteness, mood and pronoun reference or facilitate grammatical processes such as embedding, compounding and rearrangement. Specifying further the role of the various parts-of-speech is needed in order to capture more accurately the intent of a query and therefore to render the query enhancement process more effective.

The concept-role highlighting procedure relies on two subsequent steps: (i) Grammatical relations linking query constituents are determined, *head* (i.e. key phrasal concept) and *dependents* (i.e. all terms linked to the *head*) are determined; (ii) Concept roles are assigned according to the functionality of the constituents related by the grammatical relations. These relations are defined to imply the appropriate role assignment for query concepts (Selvaretnam & Belkhatir 2016).

In the Statistical Processing step (Module 1b), frequently co-occuring terms are first produced by matching the formed grammatically dependent base pairs against word sequences of *n*-windows size from an n-gram model. The n-gram matches are sorted in descending frequency order of occurrence and subsequently the terms surrounding the base pair are extracted as statistically collocated expansion terms. The frequencies of the candidate terms are obtained from a *1*-gram model. All candidate terms are tokenized and the frequencies of terms sharing the same root word are summed prior to sorting to avoid multiple morphological variants to be ranked at the top. A sample of statistically-collocated candidate expansion terms extracted based on the base pairs "*overcrowded ↔ prisons*" and "*coping ↔ prisons*" for the example query (i.e. *state, years, jails, country, conditions, problems*) are shown in fig. 4.

After tagging the original query terms with a role-type, lexical-semantic related candidate expansion terms are only extracted for query terms which are labelled with the role-type *CoI* and *Dc* as they represent the intent of the search goal. The main relations that are considered in this publication are the lexically-related synonyms as well as semantic relations i.e. hypernyms, hyponyms and coordinate terms. For the process of extraction to take place without flaw, accurate senses which had been assigned through the disambiguation process *in Step 4* are utilized. The average semantic similarity between a candidate expansion term and all significant original query terms

(*base terms* annotated with role-type *CoI* and *Dc*) is calculated in order to rank the candidate terms in order of relevance to the entire query. A sample of synonymously related candidate expansion terms extracted based on the base terms "*overcrowded*", "*prisons*" and "*coping*" for the example query i.e. *prison house, grapple, deal, contend* and *make out* are shown in fig. 4.

## 5.2 Cognitive Pattern-Based Reconciliation

Reformulation of queries happen when users are dissatisfied with the results of their search or they want to refocus their search based on the results obtained at the first attempt. Although there may be varying user behaviours in the search process, several reformulation patterns can typically be observed from search logs. Three major reformulation possibilities e.g. resource, format and content modifications were identified by Rieh and Xie (06) with content modifications being the most likely aspect to be emulated for automated query enhancement. Content modification refers to the changes made by replacement with synonyms, as well as amendments resulting in generalization, specialization, and parallel movement of the query. The results of the study reveal that users do not always start with a general query and attempt to specify it. In fact, parallel movement is the most popular means of content modification (51.4%) while specification accounts for 29.1% and generalization for 15.8% of all modified queries. Users do not often change a query by simply replacing it with a synonym i.e. only 4.9% of all expansions are related to patterns of replacement with a synonym. These content modifications require the removal of original query terms prior to including new related terms. However, when employing automated query expansion, it would not be advisable to remove any term used in the formulation of the original query so as to ensure that the user's intent is always maintained. As such, within this framework, these reformulation strategies are revised to enhance an existing query through the addition of new related expansion terms.

The query reformulation patterns are then redefined in order to reflect these changes, as follows:
- **Generalization** entails the addition of broader terms which widens the search scope related to search goal.
- **Specialization** entails the addition of narrower terms which refocuses the query to a more particular topic related to the search goal.
- **Enhancement with synonyms** entails the addition of terms which share the same meaning and as such employ all vocabulary that are similar to the search goal.
- **Parallel movement** entails the addition of terms which are closely related to a search goal by means of belonging to a different aspect of the same topic.

These query reformulation patterns only include certain lexical-semantic related terms in the final expanded query. However, the inclusion of *both* statistically-collocated and lexical-semantically related terms is expected to improve retrieval performance. Statistically-collocated terms are assumed to be related to a particular original query term due to its frequency of use within the same context. As such, a final set of four integration expansion patterns (IE) are defined for the query expansion process within this framework. The statistically-collocated terms extracted are included alongside the respective lexical-semantic related terms needed according to the four content modification strategies:

***Integrated Expansion 1 (IE1)*** *is defined as the process of expanding an original query with statistically-collocated expansion terms as well as synonymous terms which achieve the enhancement with synonyms content modification strategy.*

***Integrated Expansion 2 (IE2)*** *is defined as the process of expanding an original query with statistically-collocated expansion terms as well as broader terms which achieve the query generalization content modification strategy.*

***Integrated Expansion 3 (IE3)*** *is defined as the process of expanding an original query with statistically-collocated expansion terms as well as specific terms which achieve the query specialization content modification strategy.*

***Integrated Expansion 4 (IE4)*** *is defined as the process of expanding an original query with statistically-collocated expansion terms as well as terms which achieve the parallel movement content modification strategy.*

All the IE patterns are achieved through the expansion of the original query with the terms extracted through the statistically-collocated term pooling process expressed in Section 5.2.1. The top n expansion terms to be included in a final expanded query are selected from the global pool of statistically-collocated terms. The terms in the global pool are ranked based on their frequency of occurrence and the cut-off threshold for term selection is experimentally determined. In the event that several terms share the same frequency of co-occurrence, the expansion terms are selected randomly.

Then, each integrated query expansion pattern defined requires the inclusion of one particular lexical-semantic relation and has varying impact on an original query. In the case of *IE1*, the pool consists of several groups of synonyms which are generated for each base term within a query. In the case of *IE2* and *IE3*, the pool of extracted terms consists of several levels of hypernyms and hyponyms representing increasingly specific or increasingly general terms as a result of ascending or descending traversal of the ontological hierarchy, respectively. On the other hand, for the *IE4* pattern, all returned candidate terms for the related-To association are from the same level of the hierarchy. For all IE patterns, if the candidate terms have other terms that share the same meaning, its corresponding synonyms are also included in the pool. The global relatedness of each candidate expansion term to all base terms in an original query is computed and the terms within each global pool (i.e. synonyms pool, hypernyms pool, hyponyms pool, coordinate terms pool) are ranked based on their degree of relatedness. For each IE pattern, the corresponding global pool of terms is referred and the cut-off threshold for selecting the terms to be included in a final expanded query is experimentally determined.

The selection of terms to be included in the final expanded query is done independently according to the category of terms it belongs to, for each defined IE pattern (*IE1*, *IE2*, *IE3* and *IE4*). Candidate terms which may have a detrimental effect (e.g. special characters, common words etc) on retrieval performance are removed in the respective term pooling processes. It is, however, necessary to eliminate overlaps or semantically-equivalent terms between the selected set of expansion terms from the statistically-collated and lexical-semantic related term pools to ensure that there are no repetitions in the final expanded query. As such, approximate string matching is applied to remove multiple occurrences of identical terms in the final selected set of terms. A pair of terms is deemed semantically equivalent if they are identical, either directly or through their lexical unit e.g. "*book*" or "*books*" are deemed equivalent. A single occurrence of the term will be retained as a lexical-semantic related expansion term as this category of expansion terms is assumed to have a higher degree of importance over statistically-collocated terms. This assumption is made on the basis that lexical-semantic expansion terms are extracted from a formalized lexical knowledge-base that defines specific notions of relations which reflect meaning through linguistic associations between terms. For example, if a co-occurring candidate term e.g. "*journal*" matches the hyponym of an original query term "*book*", then the term "*journal*" would be retained as a hyponym of book and assigned the weight set for hyponyms and removed from the list of statistically-collocated candidate expansion terms. The number of optimal terms as well as system parameters that would optimize the retrieval results vary considerably, depending on the query and the retrieval model employed. As such the optimal number of terms to select for improved retrieval effectiveness is experimentally determined. These terms (i.e. original and expansion terms) have varying significance within the context of a query and as such coupled with a GA-based weighting scheme (presented in Section 4.2 and algorithmically instantiated in Section 6.3) in order to appropriately assess their importance.

# 6. ALGORITHMIC INSTANTIATION
## 6.1 Linguistic Processing
The algorithm below summarizes the implementation of the concept-based linguistic query processing module:

```
Input: Q ← The set of queries
1: for all q in Q do
2:    List queryConcepts=segmentIntoConcepts()
3:    List NcPList = getNcPList()
4:    if q is in NcPList
5:       Edit query to identify NcPs
6:    List RCP = getRelations&ConceptPairs()
7:    for all rcp in RCP do
8:       Assign role-type to each concept in pairs
9:    for all c in queryConcepts do
10:      List RACP=getRelations&AssignedConceptPairs()
11:      for all racp in RACP do
12:         List ConceptAssignments = getConceptAssignments()
13:         for all ca in ConceptAssignments do
14:            if ca is left unassigned
15:               Resolve according to two cases
16:                  Case 1: Frequency dependence
17:                  Case 2: Inheritance dependence
18:            if ca has multiple assignments
19:               Resolve based on Priority Relation
20:         end for
21:      end for
22:   end for
23:   end for
24: end for
```

An ontology of NcPs (*NcPList*), constructed from several sources such as ontologies (e.g. WordNet) and standard lists found on the Web, is used to identify NcPs in the input queries.

Concept pairs of the parse output gathered in the list *RCP* are assigned roles based on the generated role types. Concepts left unassigned (in this case, the most generic grammatical relation is generated) or those having multiple assignments are singled out and processed according to two techniques. The *Inheritance dependence* technique considers that if a concept left unassigned is assigned a role in one of the other derived relations, then this role is automatically propagated. If it is not the case, then the *Frequency dependence* technique considers two heuristics: (i) the concept occurring more frequently is assigned the *CoI* role while the second is tagged as a *DC*; (ii) if both concepts have equal frequency, they are considered *CoIs*. The case of concepts with multiple assignments occurs owing to the essence of natural language queries in which non-adjacent and adjacent concepts are semantically or syntactically linked. The *Priority Relation* rule is proposed to address concepts with multiple assignments in the *ConceptAssignments* structure where the more noteworthy role is kept.

## 6.2 Integrated Enhancement
The integrated enhancement process is represented in the algorithm below:

```
Input: Q ← The set of web queries
1: for all q in Q do
2:    Select candidate terms from ranked statistically-collocated term pool
3:    If Integrated Expansion Pattern is IE1
4:       Select terms from ranked synonyms pool
5:    If Integrated Expansion Pattern is IE2
6:       Select terms from ranked hypernyms pool
7:    If Integrated Expansion Pattern is IE3
8:       Select terms from ranked hyponyms pool
9:    If Integrated Expansion Pattern is IE4
10:      Select terms from ranked coordinate terms pool
```

11:   For each candidate term in each selected term set
12:     Identify duplicate occurrences of identical terms across term pool
13:     Retain single entry in lexical-semantic related term pool
16: end for
22: end for

In accordance to an experimentally determined optimal number of expansion terms, expansion terms are selected from the statistically-collocated and lexical-semantic related global term pools. The selection of lexical-semantic related terms is dependent on the integrated expansion pattern that is being emulated. For *IE1*, *IE2*, *IE3* and *IE4*, the terms are selected from the synonym, hypernym, hyponym and coordinate term pools respectively. Since the two global pools of candidate terms had been formed in two separate term pooling processes, the possibility of overlap in the selected terms exists. As such redundant terms are identified and their occurrences are retained as part of the lexical-semantic related term expansion set.

## 6.3 Concept Weighting

The algorithm below describes the implementation of the concept weighting module:

```
Input: Q ← The set of Web queries
1: for all q in Q do
2:    InitializingGeneticAlgorithmParameters();
3:    String [ ] roleWeights = extractWeights()
4:    while (iterationNumber != maxNumberIterations) do
5:       Build input file for retrieval with generated weights
6:       generatedMAP = getMAPfromToolkitOutputFile();
7:       if (generatedMAP > maxMAP)
8:          Replace maxMAP value with generatedMAP value
9: end for
```

The enhanced query is then subjected to the application of role-type weights on its terms. Considering the implementation of a GA to maximize the MAP, a population of chromosomes characterizing the produced weights (e.g. *CoI, Dc, Rc, Sc, Ec*) is generated. These are then predicted and optimized starting with the *InitializationOfGAParameters* method and the application of crossover and mutation genetic operators which are set at a rate of 1000 and 10 respectively. Role weights are determined for four out of the five role types (i.e. *CoI, Rc, Dc* and *Ec*) furthermore setting a restriction of values within the interval [0,1]. In order to eliminate bias on general concepts, the role type *Sc* is assigned a null weight. A chromosomal population of 200 entities characterizing the role-type weights is evolved over 100 iterations and serves as a termination condition. The generated list of weights (i.e. *roleWeights*) is used as an input for the retrieval process. Upon obtaining the retrieval results, *tempMAP* is assigned the MAP value of the query extracted from the output file. If the *generatedMAP* value is greater than the *maxMAP* obtained in preceding iterations, then the latter is updated. Each chromosome's fitness value benefits from a boost according to the global MAP value with the biggest boost attributed to the highest MAP (i.e. above 0.5) across all iterations. Fitness value is moreover employed to assess the number of iterations as the termination condition to control the processing load. GA runs are launched for all processed queries.

## 7. EVALUATION

Herein, we present the test collections, introduce the evaluation measures, describe the compared frameworks then present the retrieval results. We conclude with a discussion of the obtained empirical results.

---

<top>

<num> Number: 151

<title> Topic: Coping with overcrowded prisons

<desc> Description:

The document will provide information on jail and prison overcrowding and how inmates are forced to cope with those conditions; or it will reveal plans to relieve the overcrowded condition.

<narr> Narrative:

A relevant document will describe scenes of overcrowding that have become all too common in jails and prisons around the country. The document will identify how inmates are forced to cope with those overcrowded conditions, and/or what the Correctional System is doing, or planning to do, to alleviate the crowded condition.

</top>

---

**Figure 5: Sample TREC Topic "coping with overcrowded prisons"**

## 7.1 Test Collections

The proposed solution is evaluated on the ad hoc collections of the Text REtrieval Conference (i.e. Tipster Disk 1&2 comprising Wall Street Journal (WSJ) documents, Associated Press (AP), Federal Register (FR), Department of Energy (DOE) and Computer Select as well as GOV2 (i.e. .gov sites)). These collections were compiled as part of a research evaluation campaign by the National Institute of Standards and Technology (NIST) and range in size between 74, 520 documents to 25,114,919 documents. As far as evaluation is concerned, the datasets are sorted relative to size and grouped into small clusters (i.e. with up to 100, 000 documents), medium clusters (i.e. consisting of above 100, 000 documents and below 250,000 documents) and large clusters (i.e. with more than 250, 000 documents). This dataset segmentation is performed based on the average size of Tipster datasets (i.e. 205,785 documents approximately) and the largest dataset (i.e. GOV2) employed herein. The test topics comprise *title* and *description* fields which characterize search needs (cf. Fig. 5). In order to build a query collection with varying query lengths and applicability to heterogeneous collection sizes, we examine the *title* field of the fifty test topics formulated in the TREC 1, 3, 8 and Terabyte evaluation suites (cf. Table 5). Binary judgments assessing document relevance are associated to each topic and used in the quantification of search performance which is based on the number of relevant documents that are retrieved successfully.

**Table 5: Summary of TREC Collections & Topics**

| Dataset | # Docs | Topics | TREC |
|---|---|---|---|
| WSJ90_92 | 74,520 | 51-100 | 1 |
| AP88-90 | 242,918 | 51-100 | 1 |
| SJM1991 | 90,257 | 51-100 | 1 |
| WSJ87_92 | 173,252 | 151-200 | 3 |
| AP88_89 | 164,597 | 151-200 | 3 |
| Disk4&5 | 489,164 | 401-450 | 8 |
| GOV2 | 25,114,919 | 751-800 | Terabyte |

## 7.3 Retrieval Evaluation

To assess the performance of the proposed framework, we compare it against :

i) Unigram Language Model (LM) (Ponte & Croft, 1998), its implementation produces the baseline results of the original queries without any modification.

ii) Relevance Model (RM) (Lavrenko & Croft, 2001), its performance translates the baseline performance of the state-of-the-art pseudo relevance feedback technique taking into account the top n frequently occurring terms extracted from the top k documents resulting from the search.

iii) Concept-Role Mapping (CRM) enhancement where query constituents are annotated in accordance to the concept-role mapping process and provided a weight reflective of the role-type in representing the informational intent (Selvaretnam&Belkhatir 2016). This approach is representative of frameworks performing the processing steps described in Module 1 of Section 2.

iv) Linguistic-Statistical (LSTAT) enhancement framework where base pairs are formed based on the notion of term dependencies (Huston & Croft 14), specifically sequential dependence (i.e. dependence is assumed to exist between adjacent query terms). This approach is representative of frameworks coupling the processing steps described in Modules 1 & 2 of Section 2.

v) Linguistic-Semantic (LSMT) enhancement framework integrating linguistic and extrinsic knowledge-based processing (Selvaretnam&Belkhatir 2019). It is declined in four variants performing enhancement with : synonyms (LSMT_Syn), hypernyms (LSMT_Hyper), hyponyms (LSMT_HYPO) and coordinate terms (LSMT_COORD). This approach is representative of frameworks coupling the processing steps described in Modules 1 & 3 of Section 2.

Our empirical Integrated Linguistic/Statistical/Semantic (ILSS) framework is itself declined into four experimental variants: ILSS_IE1, ILSS_IE2, ILSS_IE3 and ILSS_IE4, corresponding to the process of linguistic pre-processing of queries, candidate expansion term pooling of both statistically-collocated and lexical-semantic related terms (synonyms for ILSS_IE1, hypernyms for ILSS_IE2, hyponyms for ILSS_IE3 and coordinate terms for ILSS_IE4) as well as the pattern-based reconciliation of candidate terms from multiple knowledge sources. Enhancement concepts weighted according to the GA process are further adjoined to the initial query concepts.

Regarding the evaluation metric, Mean Average Precision (MAP) is used to assess the effectiveness of the compared systems similar to related research. The paired t-test with 95% confidence level ($p < 0.05$) is used to assess the statistical significance of differences in terms of MAP of the compared frameworks. In the remainder (i.e. Tables 6, 7, 8 and 9), superscripts $\alpha$, $\beta$, $\delta$, $\gamma$ and $\mu$ indicate statistically significant improvements in MAP over five variations, i.e. LM, RM, CRM, LSTAT and LSMT respectively.

In the discussion of Section 7.5, we furthermore compare our improvements in terms of MAP with respect to those obtained by Collins-Thompson&Callan (2015) and Dipasree et al. (2014) which are comparable in terms of proposed frameworks and experimental datasets. We also compare our results to those obtained by Roy et al. (2016) who used word embeddings for query expansion. The reason for including these results in the discussion section is explained by the fact that their reported baseline MAP results differ from those shown herein, which is often the case even on identical test sets as reported by Carpineto & Romano (2012). Accordingly, we will therefore compare the comparable increases in MAP obtained over baseline frameworks.

## 7.4 Results

### 7.4.1 Integrated enhancement with synonymy pattern

**Table 6: Performance of ILSS_IE1 in Retrieval Tasks**

| Query No | Dataset | LM (MAP) | RM (MAP) | CRM (MAP) | LSTAT (MAP) | LSMT_Syn (MAP) | ILSS_IE1 (MAP) |
|---|---|---|---|---|---|---|---|
| 51-100 | WSJ90-92 | 0.1874 | 0.2011 | 0.2178 | 0.2182 | 0.2299 | 0.236 $^{\alpha\beta\delta\gamma\mu}$ |
| | AP88-90 | 0.1979 | 0.2482 | 0.2314 | 0.249 | 0.2558 | 0.262 $^{\alpha\beta\delta\gamma\mu}$ |
| | SJM1991 | 0.1463 | 0.1658 | 0.1741 | 0.1831 | 0.1915 | 0.2001 $^{\alpha\beta\delta\gamma\mu}$ |
| 151-200 | WSJ87-92 | 0.2352 | 0.2874 | 0.2873 | 0.3016 | 0.3357 | 0.357 $^{\alpha\beta\delta\gamma\mu}$ |
| | AP88-89 | 0.2575 | 0.3252 | 0.3063 | 0.3265 | 0.3558 | 0.3644 $^{\alpha\beta\delta\gamma\mu}$ |
| 401-450 | Disk 4-5 | 0.1926 | 0.2149 | 0.2124 | 0.2193 | 0.2281 | 0.2522 $^{\alpha\beta\delta\gamma\mu}$ |
| 751-800 | GOV2 | 0.2944 | 0.3063 | 0.3221 | 0.3284 | 0.34 | 0.3488 $^{\alpha\beta\delta\gamma\mu}$ |

The ILSS_IE1 experimental results are tabulated in Table 6 and the compilation of relative improvements shown in Figure 6. Statistically significant improvement in retrieval is observed across all variations and datasets. The improvements obtained over LM and RM range from 18.5% to 51.8% and 5.6% to 24.2%, respectively. This integrated approach also surpasses the performance of the CRM approach of role-type mapping in the retrieval process within the range of 8.3% to 24.3%. The added retrieval effectiveness achieved through ILSS_IE1 in comparison to the LSTAT techniques ranges from 5.2% to 18.4%. On the other hand, the percentage of improvements in MAP attained comparative to LSMT_Syn varies from 2.4% to 10.6%.

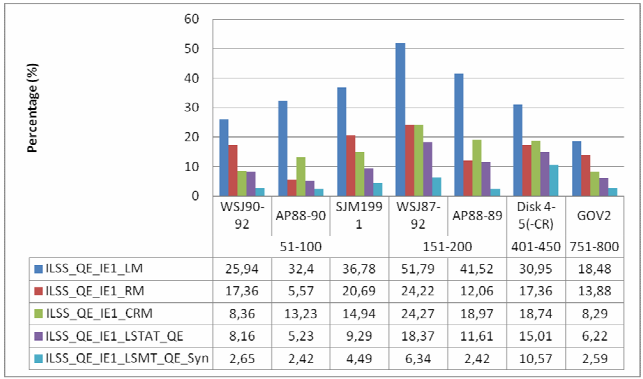

**Figure 6: Relative Performance of the ILSS_IE1 Pattern**

*7.4.2 Integrated enhancement with hypernymy pattern*
The obtained retrieval results for the ILSS_IE2 experimental variation are presented in Table 7 and relative changes in retrieval performance reflected in Figure 7.

**Table 7: Performance of ILSS_IE2 in Retrieval Tasks**

| Query No | Dataset | LM (MAP) | RM (MAP) | CRM (MAP) | LSTAT (MAP) | LSMT_Hyper (MAP) | ILSS_IE2 (MAP) |
|---|---|---|---|---|---|---|---|
| 51-100 | WSJ90-92 | 0.1874 | 0.2011 | 0.2178 | 0.2182 | 0.2184 | 0.219$^{αβδγμ}$ |
|  | AP88-90 | 0.1979 | 0.2482 | 0.2314 | 0.249 | 0.255 | 0.2638$^{αβδγμ}$ |
|  | SJM1991 | 0.1463 | 0.1658 | 0.1741 | 0.1831 | 0.1853 | 0.2$^{αβδγμ}$ |
| 151-200 | WSJ87-92 | 0.2352 | 0.2874 | 0.2873 | 0.3016 | 0.3404 | 0.341$^{αβδγμ}$ |
|  | AP88-89 | 0.2575 | 0.3252 | 0.3063 | 0.3265 | 0.3311 | 0.342$^{αβδγμ}$ |
| 401-450 | Disk 4-5 | 0.1926 | 0.2149 | 0.2124 | 0.2193 | 0.222 | 0.223$^{αβδγμ}$ |
| 751-800 | GOV2 | 0.2944 | 0.3063 | 0.3221 | 0.3284 | 0.3304 | 0.333$^{αβδγμ}$ |

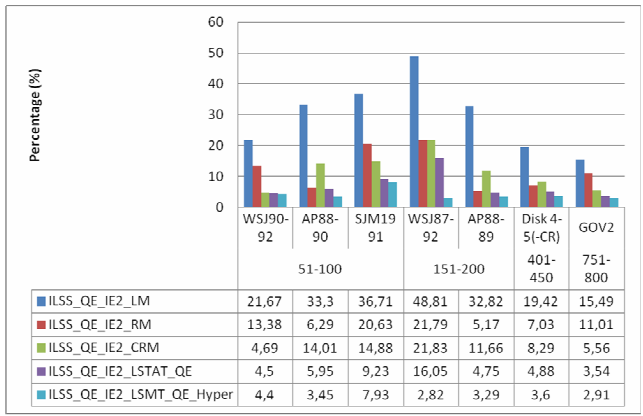

**Figure 7: Relative Performance of the ILSS_IE2 Pattern**

The ILSS_IE2 enhancement approach shows large improvements over LM (i.e. ranging from 15.5% to 48.8%) and RM (i.e. ranging from 5.2% to 21.8%). The ILSS _IE2 method which builds on the CRM approach results in increments in MAP ranging from 4.7% to 21.8%, reiterating the benefit of adopting an integrated enhancement approach. Significant improvements in retrieval performance through ILSS_IE2 are observed in comparison to separately incorporating statistically-collocated terms (LSTAT) and lexical-semantic related terms (LSMT_Hyper). Increases in MAP ranging from 3.5% to 16.1% are achieved through the integrated approach comparative to LSTAT. Similar progress was attained through the ILSS_IE2 approach compared to LSMT_Hyper where the range of improvements seen are from 2.8% to 7.9%. These significant improvements in MAP achieved across all datasets and experimental variations validate the use of an integrated approach of combining hypernyms and statistically-collocated terms in the expansion process.

*7.4.3 Integrated enhancement with hyponymy pattern*
Retrieval performance achieved through the application of the ILSS_IE3 enhancement pattern in the final query is shown in Table 8.

**Table 8: Performance of ILSS_IE3 in Retrieval Tasks**

| Query No | Dataset | LM (MAP) | RM (MAP) | CRM (MAP) | LSTAT (MAP) | LSMT_Hypo (MAP) | ILSS_IE3 (MAP) |
|---|---|---|---|---|---|---|---|
| 51-100 | WSJ90-92 | 0.1874 | 0.2011 | 0.2178 | 0.2182 | 0.2164 | 0.1899$^{α}$ |
|  | AP88-90 | 0.1979 | 0.2482 | 0.2314 | 0.249 | 0.2322 | 0.2329$^{α}$ |
|  | SJM1991 | 0.1463 | 0.1658 | 0.1741 | 0.1831 | 0.1778 | 0.1815$^{αβδμ}$ |
| 151-200 | WSJ87-92 | 0.2352 | 0.2874 | 0.2873 | 0.3016 | 0.328 | 0.3001$^{αβδ}$ |
|  | AP88-89 | 0.2575 | 0.3252 | 0.3063 | 0.3265 | 0.3125 | 0.3204$^{αδμ}$ |
| 401-450 | Disk 4-5 | 0.1926 | 0.2149 | 0.2124 | 0.2193 | 0.215 | 0.2152$^{α}$ |
| 751-800 | GOV2 | 0.2944 | 0.3063 | 0.3221 | 0.3284 | 0.3145 | 0.3272$^{αβμ}$ |

Relative improvements in MAP over the baselines LM and RM as well as CRM, LSTAT and LSMT_Hypo are indicated in Figure 8.

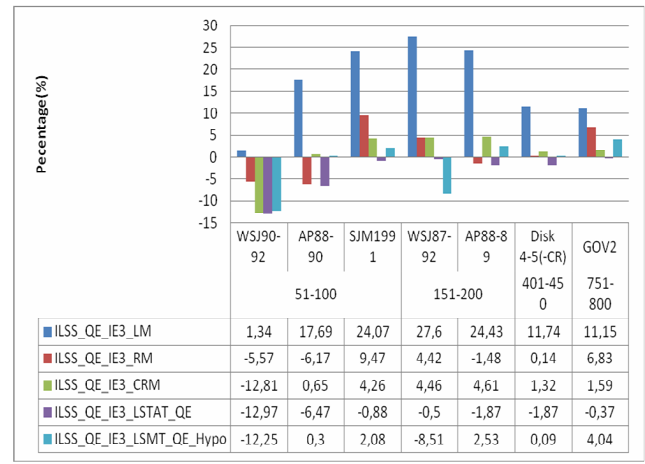

**Figure 8: Relative Performance of the ILSS_IE3 Pattern**

A combination of improvements and deteriorations is observed from the experimental tests for ILSS_IE3. A significant amount of improvement is seen over the baseline LM approach. On the other hand, when compared against the RM expansion method, both positive (i.e. an increment in MAP ranging from 4.4% to 9.5%) and negative (i.e. a decrement in MAP ranging from 5.6% to 6.2%) impacts are observed. A similar pattern of results is also noticed compared to the CRM and LSTAT approaches. Although an average of 4.4% of increase in MAP is observed on a subset of the datasets examined, decrease in MAP is also noticed compared to CRM. Implementation of ILSS_IE3 gives a negative impact on retrieval compared to LSTAT where all variations reveal a reduction in MAP. Unlike the generally better performance of LSTAT compared to ILSS_IE3, the integrated approach exhibits significant improvements over LSMT_Hypo where an increase in MAP from 2.1% to 4.0% is perceived. However, reduction in MAP is also evident over LSMT_Hypo within the range of -8.5% to -12.3%. This decrease in retrieval performance over LSMT_Hypo is due to the existence of statistically-collocated terms within the expanded query as well as multiple hyponyms for multiple base terms causing general diversification of the query.

*5.4.4 Integrated Enhancement with parallel movement*
Experimental results obtained after applying the parallel movement strategy are shown in Table 9 and the compilation of relative improvements displayed in Figure 9.

**Table 9: Performance of ILSS_IE4 in Retrieval Tasks**

| Query No | Dataset | LM (MAP) | RM (MAP) | CRM (MAP) | LSTAT (MAP) | LSMT_CoordTerms (MAP) | ILSS_IE4 (MAP) |
|---|---|---|---|---|---|---|---|
| 51-100 | WSJ90-92 | 0.1874 | 0.2011 | 0.2178 | 0.2182 | 0.2171 | 0.1899$^{\alpha\mu}$ |
|  | AP88-90 | 0.1979 | 0.2482 | 0.2314 | 0.249 | 0.2346 | 0.1979$^{\mu}$ |
|  | SJM1991 | 0.1463 | 0.1658 | 0.1741 | 0.1831 | 0.1751 | 0.1831$^{\alpha\beta\delta}$ |
| 151-200 | WSJ87-92 | 0.2352 | 0.2874 | 0.2873 | 0.3016 | 0.3327 | 0.2958$^{\alpha\beta\delta\mu}$ |
|  | AP88-89 | 0.2575 | 0.3252 | 0.3063 | 0.3265 | 0.3097 | 0.3169$^{\alpha\delta}$ |
| 401-450 | Disk 4-5 | 0.1926 | 0.2149 | 0.2124 | 0.2193 | 0.2145 | 0.213$^{\alpha}$ |
| 751-800 | GOV2 | 0.2944 | 0.3063 | 0.3221 | 0.3284 | 0.329 | 0.3338$^{\delta}$ |

The parallel movement pattern also returns increased retrieval effectiveness compared to LM as seen with other integrated expansion approaches. It shows a larger range of deterioration (i.e. from -2.6% to -20.3%) in MAP than any of the achieved increases (i.e. from 2.9% to 10.4%) when compared to RM. Again a similar pattern is observed with the percentage of change in MAP observed over CRM. Despite having several significant increments in MAP (i.e. from 3% to 5.2%), a larger decrease in performance is seen on two datasets (i.e. -12.8% for WSJ and -14.5% for AP) as compared to CRM. ILSS_IE4 revealed statistically significant deterioration across most datasets (i.e. -3% to -20.5%) when compared against LSTAT while an insignificant increase in MAP of 1.7% was achieved on the GOV2 dataset only. In contrast, ILSS_IE4 showed lesser deterioration in MAP (i.e. -2.3% to -4.4%) over LSTAT compared to LSMT_CoordTerms. The improvements in MAP observed which range from 12.5% to 18.5% were however not statistically significant. This is indicative that the parallel movement based integration method would be difficult to automate as it is not possible to deduce a users' preferred redirection in search goal without application of extensive personalization techniques.

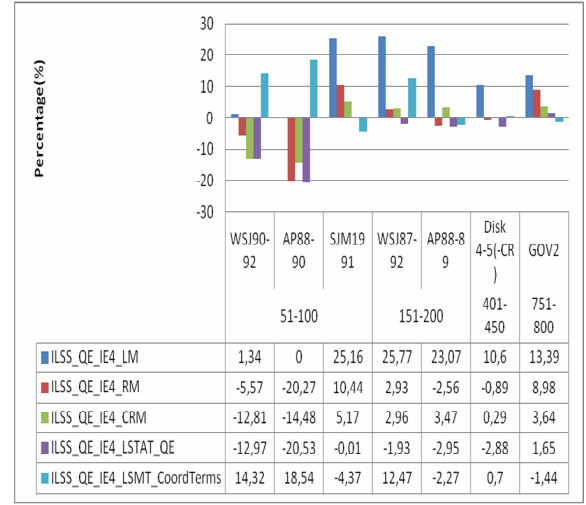

**Figure 9: Relative Performance of the ILSS_IE4 Pattern**

## 7.5 Experimental Discussion
The retrieval performances among the four IE patterns are shown in Figure 10. The integration of statistically-collocated expansion terms with lexical-semantic related terms in a final expanded query has shown marked improvements in retrieval performance over the de facto baselines as well as independent expansion. ILSS_IE1 and ILSS_IE2 revealed statistically significant improvements in MAP on most test cases compared to ILSS_IE3 and ILSS_IE4. ILSS_IE3 and ILSS_IE4 especially do not perform well when the collection size is large. This is due to the heterogeneity that exists in larger document collections which when coupled with expansion terms of differing directions in search goal (i.e. through the use hyponyms and coordinate terms) negatively impact query performance. When expanded based on ILSS_IE2, the results showed a smaller decline in MAP comparatively (i.e. between -4.5% to -11.5%) over ILSS_IE1. It also showed similar retrieval performance as ILSS_IE1 on two average sized test collections. On the other hand, ILSS_IE3 and ILSS_IE4 showed significant deterioration in MAP ranging between -4.3% to -24.5% and between -6.2% to -19.5%, respectively when compared to ILSS_IE1. These results reiterate our concerns related to the fact that in automated query enhancement it is difficult to predict which of the many hyponyms or coordinate terms may be of interest to a user. The inclusion of multiple numbers of such relations would result in conflicting redirection of the query.

We propose to compare our improvements with those obtained by Collins-Thompson & Callan (2005) making use of multiple expansion term types inclusive of synonyms, stemmed terms, associated terms as well as co-occurring terms from the Web and top retrieved documents when processing the queries, by Roy et al. (2016) who propose a word embedding-based model where terms that are related to the query terms are pooled using the K-nearest neighbor algorithm and by Dipasree et al. (2014) who use the WordNet ontology in the expansion process and.

The first report relative improvements garnered for query set 401-450 (TREC 8) on dataset Disk4&5 over RM being at best 2.3% unlike the 17.4% relative increase in MAP achieved with ILSS_IE1 and 7% achieved with ILSS_IE2. The marginal increase in retrieval performance is due to the fact that important factors have not been examined and utilized extensively: the query structure, linguistic characteristics, the term dependency model utilized in statistical processing and possible term relationships to extract. On the same query set and dataset, the second report relative improvements over LM being at best 8.2% and deteriorated performance compared to RM. These results reiterate the fact that detailed linguistic examination of queries and proper semantic analysis are crucial to derive effective enhancement procedures. Finally, Dipasree et al. (2014) publish relative improvements of approximately 16% noted over baseline RM. We shall remark that a noticeable limitation of WordNet is the limited coverage of concepts and phrases within the ontology (Maree&Belkhatir 2015). This paper hypothesizes that extrinsic language resource based query expansion would benefit most in the case where multiple sources are utilized in the concept generation process.

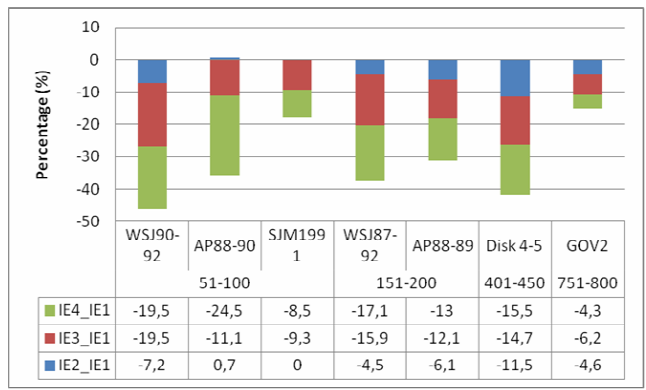

**Figure 10: Relative Performance of Integrated Enhancement**

## 8. CONCLUSION

In this paper, we proposed a query enhancement framework making use of intrinsic linguistic analysis to highlight the role of query concepts. Additionally, the use of an extrinsic knowledge based resource made it possible to spawn concepts semantically coherent with the query content. We furthermore considered the application of enhancement patterns emulating user behaviour and eventually integrating statistical and extrinsic language based resources to ensure that only closely related concepts are taken into account in the enriched query. The inclusion of various query-document links coupled with an optimization-based weighting scheme has demonstrated improvements in retrieval performance. In our future work, we will further investigate the impact of using enriched semantic resources such as proposed in (Maree&Belkhatir 2015) to improve the query enhancement process.